# The Structural Origin of Enhanced Dynamics at the Surface of a Glassy Alloy

Gang Sun, Shibu Saw, Ian Douglass and Peter Harrowell

School of Chemistry, University of Sydney, Sydney NW 2006 Australia

Abstract

The enhancement of mobility at the surface of an amorphous alloy is studied using a combination of molecular dynamic simulations and normal mode analysis of the non-uniform distribution of Debye-Waller factors. The increased mobility at the surface is found to be associated with the appearance of Arrhenius temperature dependence. We show that the transverse Debye-Waller factor exhibits a peak at the surface. Over the accessible temperature range, we find that the bulk and surface diffusion coefficients obey the same empirical relationship with the respective Debye-Waller factors. Extrapolating this relationship to lower T, we argue that the observed decrease in the constraint at the surface is sufficient to account for the experimentally observed surface enhancement of mobility.

The mobility at the surface of a glass or deeply supercooled liquid can exceed that of the bulk by more than 6 orders of magnitude as measured either directly, as an increase in mobility at the surface [1], or as a depression of the surface glass transition temperature relative to that of the bulk [2,3]. Diffusion of indomethacin at the surface of the amorphous state has been measured to be $10^7$ times that of the bulk at 5K below the glass transition temperature $T_g$ [4]. The enhanced mobility at a glass surface is significantly larger than its analogue in the surface melting of crystals [5]. The (110) surface of lead [6], for example, exhibits a liquid monolayer at roughly $0.98T_m$ ($T_m$ being the melting point). The enhanced mobility at the glass surface has a number of important consequences, some of which are already the subject of active study. Crystallization has been observed to occur rapidly across the surface of a glass while remaining unobservably slow within the bulk of the sample [7]. Enhanced mobility is associated with more efficient relaxation and, hence, the formation of very low enthalpy glasses when formed by vapour deposition [8-10]. Thin polymer films are observed to de-wet a substrate considerably faster than the expectation based on bulk mobilities [11]. Other consequences of an enhanced surface mobility have yet to be studied in detail. The stability of fabricated structures on amorphous solids, such as the nanoporous metal glasses formed by selective solvation of a species, is highly dependent on the mobility of material across the surface [12]. The stability of surface structures applies equally to the inherent surface roughness of a glass, arising from the arrest of the thermally excited capillary waves [13,14]. Whether these arrested features are determined by the bulk glass transition or the surface glass transition is question that remains unresolved.

Surface enhanced mobility has been observed in simulation studies of the free surface of glass forming liquids including polymers [15], atomic liquids [16-18] and, to a lesser degree, silica [19]. While the phenomenology is established, an explicit account of the origin of the enhanced mobility is lacking. The problem is analogous to that posed by the connection between the dynamic heterogeneities in the bulk glass former and the structure. This perspective, in which the enhanced dynamics at the free surface is treated as a macroscopic dynamic heterogeneity, has already proved to be successful in modelling a range of phenomena associated with the dynamics at the glass surface and its response to temperature changes [9]. It follows, then, that a structural account of the surface mobility of a glass may be found using an approach that has previously proved valuable for dynamic heterogeneities.

Specifically, dynamic heterogeneities arise as a consequence of spatial variations in the degree of constraint imposed on particle motion by a dense amorphous configuration. The notion of a 'degree of constraint' was rendered explicitly in a harmonic treatment of the amorphous configurations, either by the distribution of soft localised normal modes [20] or the individual particle Debye-Waller factors [21,22].

In this paper we shall apply the analogous analysis to the dynamics of particles at the free surface of a glass. This will involve determining the temperature dependence of surface mobility in a glass forming alloy and the calculation of the variation of constraint between surface and bulk, using the normal modes of the glass with interface. We simulate a glass forming alloy comprised of a binary mixture of particles interacting by Morse potentials,

$$u_{ij}(r) = \varepsilon \left[ \exp\left( -2\alpha \left( \frac{r}{\sigma_{ij}} - 1 \right) \right) - 2\exp\left( -\alpha \left( \frac{r}{\sigma_{ij}} - 1 \right) \right) \right] ,$$

where $\varepsilon = 1.0$, $\alpha = 6$, $\sigma_{AA} = 1.0$, $\sigma_{BB} = 0.816$ and $\sigma_{AB} = 0.908$. This choice of parameters results in a glass forming liquid with an equilibrium crystal structure similar to that of $MgZn_2$ [23]. On this basis, our Morse mixture resembles the Lennard-Jones mixture of Wähnstrom [24] which crystallizes into the same structure. The composition is chosen as $AB_2$, the same as that of the crystal. Length and temperature are in units of $\sigma_{AA}$ and $\varepsilon/k_B$, respectively. Time unit $\tau_0$ is $\sigma_{AA}(\varepsilon/m)^{1/2}$. The temperature dependence of the self diffusion constant for this mixture has been fitted [23] to the Vogel-Fulcher-Tammann (VFT) expression,

$$D = D_0 \exp\left( -\frac{E_a}{T - T_0} \right),$$ with $D_0 = 0.43056$, $E_a = 1.201$ and $T_0 = 0.184$. We shall use $T_o$ as a reference temperature in the following analysis. The free surface is created as follows. A liquid is equilibrated at $T = 0.54$ and constant pressure $P = 0.0$, and then subjected to an energy minimization, so generating an inherent structure. Two surfaces are then created by moving

the position of the periodic boundary normal to the z axes so as to double the simulation box length, an amount sufficient to prevent any contact between the periodic images along this axis. Each of these surfaces represents the product of an idealised cleavage of the inherent structure at T = 0 and we shall refer to such a surface *unrelaxed*. Annealed versions of this surface are generated by giving the particles random velocities from a Boltzmann distribution at the desired annealing temperature and then running the trajectory for a time interval of 800 $\tau_0$ ($10^3$ atomic vibration periods). The final configuration is subjected to an energy minimization to generate a new amorphous solid surface at T = 0, but characteristic of the annealing temperature. All results presented are based on the average taken over 40 surfaces (i.e. 20 slab configurations, each with two surfaces) generated by this procedure.

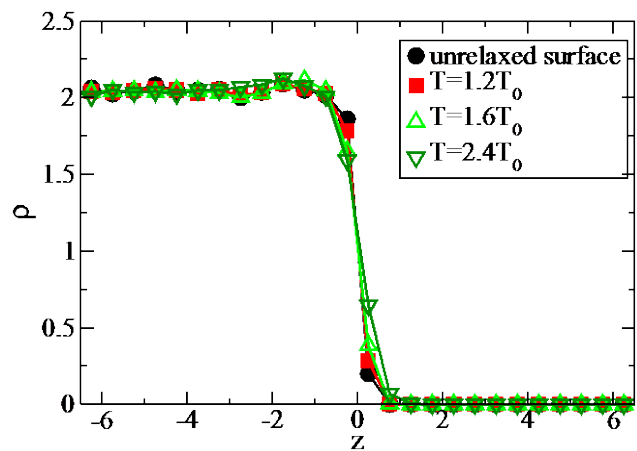

**Figure 1.** The free surface of the amorphous solids at T = 0. The interface width is ~ 2 for all surfaces. The density profiles for the unrelaxed surface and surfaces annealed at $T/T_0$ = 1.2, 1.6 and 2.4 as explained in the text.

The average density profiles of the unrelaxed and annealed surfaces are presented in Fig. 1. The interface has a width of ~ 2 large particle diameters. Only a small increase in the width is observed with increasing annealing temperature with the root mean squared variation of the position of the surface about the mean position increasing by 0.073 $\sigma_{AA}$ in going from the unrelaxed surface to the surface annealed at $2.4T_0$.

A particle is designated as a surface particle if it lies within a distance of 1.0 along the z axes from the mean position of the interface (defined here as the position of the half height of the density profile). The mean squared displacements have been measured at the various annealing temperature and the diffusion constant extracted for lateral (i.e. xy) motion within the surface particles as follows, $D_{xy} = \dfrac{1}{4} \dfrac{d\left[<\Delta x^2(t)> + <\Delta y^2(t)>\right]}{dt}\Bigg|_{<\Delta x^2>,<\Delta y^2>=1.0}$ where $<\Delta x^2>$ and $<\Delta y^2>$ are the average mean squared displacement of particles that remain in the surface region for the entire interval from $t_0$ (initial time) to t. As shown in the Fig.2 insert, the lifetime of particles in the surface region is sufficient that more than 70% of initial surface particles are still included when $<\Delta x^2>$ and $<\Delta y^2>$ equal 1.0. This diffusion constant is plotted in Fig. 2 along with the diffusion constant for the bulk liquid.

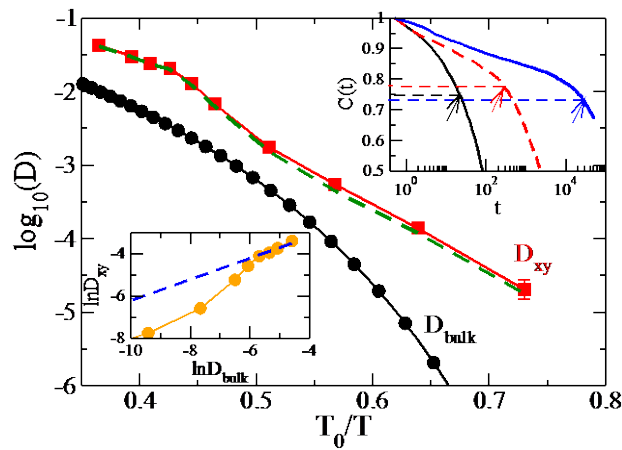

**Figure 2.** The lateral surface diffusion coefficient $D_{xy}$ (red squares) and the bulk liquid diffusion constant (calculated using the VFT expression [23]) $D_{bulk}$ (black circles) plotted as a function of $T_0/T$. The dashed line is the lateral surface diffusion constant calculated without the contribution from 'free' particles. *Insert (top right):* The fraction C(t) of surface particles

that have not left the surface after a time t for $T/T_o$ = 2.5 (black), 2.0 (red dashed) and 1.4 (blue). Arrows indicate the times marking the onset of diffusive motion. *Insert (bottom left):* $\ln(D_{xy})$ plotted against $\ln(D_{bulk})$ with the dashed line representing the suggested [25] relation, $D_{xy} \propto \sqrt{D_{bulk}}$ .

The enhanced mobility at the surface is clearly evident, and the glass transition in the surface $D_{xy}$ is depressed below the bulk value. We note that where the T dependence of $D_{bulk}$ is clearly non-Arrhenius, the lateral surface diffusion is, at low T, essentially Arrhenius. This result is consistent with previous simulations [26] and experiments [27] of the transition from Arrhenius to non-Arrhenius behaviour as one moves in from the surface to the bulk of a supercooled liquid.

The simplest explanation for enhanced mobility at a surface would be that particles occasionally extract themselves from the glass to be physisorbed at the surface and so free to move rapidly across the surface until they are re-immersed into the dense phase. This 'skater' mechanism is characterised by rapid motion that is only exhibited by a small subset of particles at any time and that these fast particles have an anomalously small number of nearest neighbours. To test this account, we have identified those surface particles at the surface annealed at $2.4T_0$ with less than 8 neighbours (there being ~15 neighbours for bulk particles). We find that that these 'free' particles make up ~7% of the surface particles. The contribution of these physisorbed particles to the surface mobility turns out to be insignificant, as can be seen from the surface diffusion coefficient calculated without the contribution from these 'free' particles plotted in Fig. 2 (dashed line). We conclude that the enhanced mobility at the surface is a collective effect, attributable to the entire surface regions, rather than the result of a handful of 'skaters'. It has previously been demonstrated that the small amplitude fluctuations of each particle position, the individual Debye-Waller factors, provide a useful measure of this degree of constraint in the bulk glass [21]. The normal modes of the inherent structures represent the natural collective description of these small amplitude motions. For each of the 20 slab configurations we have calculated the normal modes by diagonalizing the Hessian matrix of the inherent structure. The total and surface dispersion of modes, averaged over the different configurations, are plotted in Fig. 3.

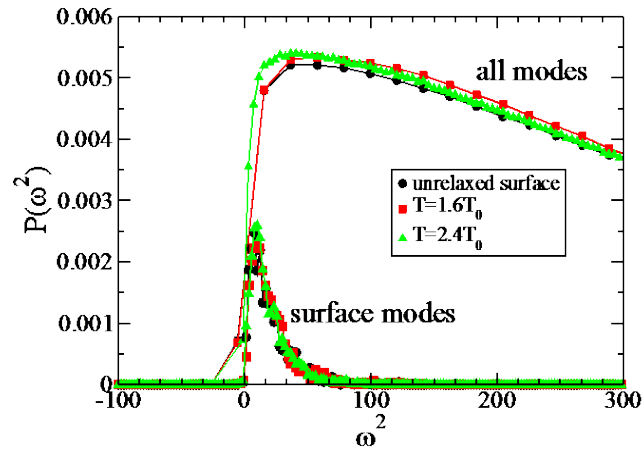

**Figure 3.** The normalized distribution of normal mode as a function of frequency $\omega$ for the amorphous slab with two free surfaces. The average number of surface modes (defined as those modes whose surface participation fraction $f > 0.2$) are also plotted. A small number (~4) of surface modes are unstable (i.e. $\omega^2 < 0$) and persist after annealing.

The participation fraction $f$ of mode $\alpha$ is defined by $f_\alpha = \dfrac{\sum\limits_{j}^{N_s} \omega_{\alpha,j}^2}{\sum\limits_{j}^{N} \omega_{\alpha,j}^2}$ where $\omega_{\alpha,j}^2$ is the eigenvalue for mode $\alpha$ and contributed by particle j, $N_s$ is the number of surface particles, and N is the total number of the glass particles. Surface modes are defined as those modes with a participation fraction $f$ of surface particles (as defined above) greater than 20%. As shown in Fig. 3, the surface modes occupy the low frequency limit of the dispersion. As defined, the surface modes have a considerable overlap with particles in the bulk glass, constituting a penetration of the surface modes into the interior of the glass sample. We have found the penetration length to be ~ 6.5, independent of the degree of annealing. Sussman et al [28] have reported that an analogous penetration length diverges as the jamming tradition is approached from the high density side for the surface of a jammed sphere packing.

The Debye-Waller factor of particle $i$ is defined as $<<(r_i(t)-r_i(t_0))^2>_t>_{isoconf}$ where the time $t_o$ corresponds to the midpoint of the plateau in the mean squared displacement and the iso-configurational average refers to the averaging over initial momenta [29]. In the harmonic approximation, the Debye-Waller factor for particle $i$ is given by $k_B T \Delta_i^2$ [22], where

$$\Delta_i^2 = \sum_k \left| \vec{v}_k^i \right|^2 / \omega_k^2 \qquad (1)$$

where $\vec{v}_k$ and $\omega_k^2$ are the $k$-th eigenvector and eigenvalue, respectively. Note that $\{\Delta_i^2\}$ are obtained directly from a configuration without the use of dynamics. We shall resolve $\Delta_i^2$ into its normal and transverse components, $\Delta_{i,z}^2$ and $\Delta_{i,xy}^2$, respectively, and plot each quantity (see Fig. 4) as a function of the particle position along the surface normal. We find that the average transverse component $<\Delta_{xy}^2(z)>$ exhibits a sharp peak at the interface, with a value ~ 3 times that of the bulk, while the average normal component $<\Delta_z^2(z)>$ only features a weaker bump (~ 1.4 times the bulk value). The decrease in constraint at the surface, as measured by the increase in the mean $<\Delta_{xy}^2(z)>$, occurs with an associated increase in the width of the distribution of individual particle amplitudes $\Lambda_{xy}^2(i)$, as shown in the Fig. 4 insert.

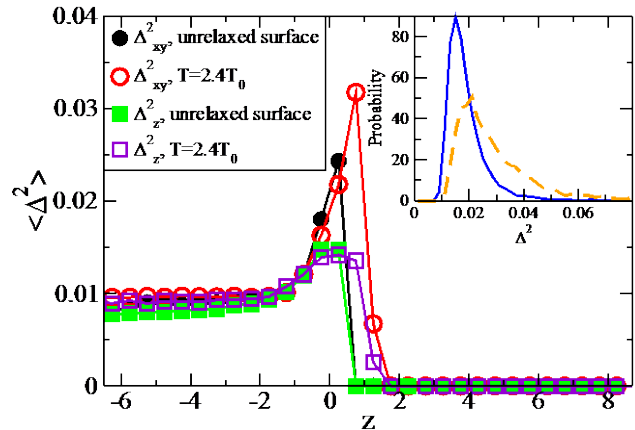

**Figure 4.** The average individual particle Debye-Waller lateral and normal factors $<\Delta_{xy}^2(z)>$ and $<\Delta_z^2(z)>$, respectively, for the unrelaxed surface and for the surface annealed at $T = 2.4T_0$. Insert: The distribution of $\Delta^2$ for the bulk (blue) and the surface (transverse) (orange dashed line) for $T = 2.4T_0$.

To complete the argument we need to connect the Debye-Waller factors to the diffusion constants. While the theoretical connection is still a work in progress [30], empirical evidence for a correlation is well established. The spatial distribution of the fluctuations in mobility of particles, the dynamic heterogeneities of the supercooled liquid, have been found to correlate strongly with the spatial distribution of the Debye-Waller factor as calculated by short time MD simulations [21] and from the normal modes [22]. Closely related, dynamic heterogeneities have been shown to correlate with the position of localised soft modes [20,31]. A number of studies [32,33] have demonstrated the strong correlation between the structural relaxation time in supercooled liquids and the magnitude of the mean squared displacement in the plateau region. In Fig. 5 we have plotted the diffusion constants for the bulk and the transverse motion in the surface against the respective Debye-Waller factors and find that, within the standard deviation of the Debye-Waller factors, they share the same dependence. We have extended the bulk data to temperatures lower than can be accessed directly by molecular dynamics by using the Volger-Fulcher expression, i.e.

$$D = A \exp\left(-\frac{B}{T - T_0}\right),$$ fitted within the simulated range [23]. The associated mean squared displacements at these inaccessible temperatures are estimated using an upper bound (the value obtained from a configuration quenched to the desired low T and certainly out of equilibrium) and a lower bound provided by the crystal at the same temperature. We find that the relationship between D and the plateau $<\Delta r^2>$ is reasonably described by an empirical relation,

$$\ln(D) = a - b/(<\Delta r^2> - c)^\alpha \qquad (2)$$

with a = 13.9, b = 5.45, c=0.0019 and $\alpha = 0.234$, where c is the value of the plateau $<\Delta r^2>$ at the $T_o$ of the Volger-Fulcher equation. Over the accessible time scales, the dependence of $D_{surf}$ and $D_{bulk}$ on the plateau $<\Delta r^2>$ are both well described by Eq.2. This observation is an interesting result, suggesting that the Debye-Waller factor determines the value of D, irrespective of whether we are at a surface or in the bulk. To estimate the low temperature enhancement $\frac{D_{surf}}{D_{bulk}}$ we shall assume that the coincidence of $D_{bulk}$ and $D_{surf}$ in Fig. 5 extends to low T.

To establish the connection between structure and dynamics, we assume that the harmonic approximation provides a reasonable approximation to the plateau value of $<\Delta r^2>$, i.e.

$$<\Delta r^2> \approx k_B T <\Delta^2> \qquad (3)$$

Inserting Eq. 3 into Eq. 2, we arrive at the prediction $\log\left(\frac{D_{surf}}{D_{bulk}}\right)$ as a function of temperature.

Using this expression, we predict a surface enhancement of mobility over that of the bulk of 6.8±2.3 orders of magnitude at T = 1.2$T_0$ ( ~ 0.91$T_g$ based on the o-terphenyl relation $T_g$ =

1.32T$_o$ [34]). This is a similar magnitude of surface enhancement to that found experimentally in indomethacine (~ 7 orders of magnitude at 0.95T$_g$ [4]) and the Pd-based metallic glass (~5 order of enhancement at 0.95T$_g$) studied by Cao et al [35].

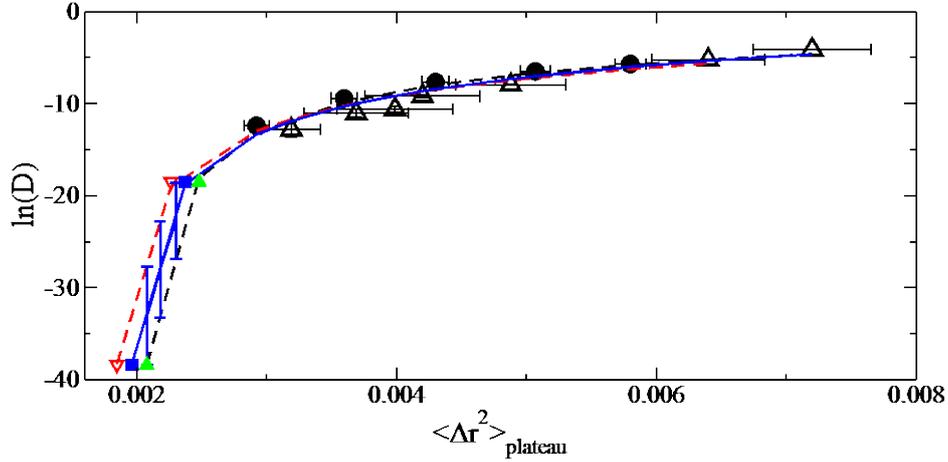

**Figure 5.** The values of D$_{bulk}$ as simulated (filled circles) or extrapolated using the VFT equation (filled squares) and D$_{xy}$ (open triangles) plotted against the plateau $\langle\Delta r^2\rangle_{plateau}$ (see text), calculated for the bulk and surface systems, respectively. Low temperature bounds on $\langle\Delta r^2\rangle_{plateau}$ are obtained from as-quenched (black dashed) and the crystal (red dashed). The solid line is the fitted expression presented in Eq. 2.

Zhang and Fakhraai [36] have shown experimentally that the surface mobility can be independent of the bulk mobility. We find a similar result. Two glasses formed from annealing at two different temperatures, 0.6T$_g$ and 1.0T$_g$, result in quite different bulk values of $\Delta^2_{bulk}$, 0.00962 and 0.01194, respectively, but very similar values of the surface term $\Delta^2_{surface}$ (i.e. 0.0136 and 0.0152). Translated via Eq. 2 into diffusion coefficients at T=0.6T$_g$, these values correspond to a difference in bulk mobility of 9±2.3 orders of magnitude while the surface mobilities lie within a factor of 5 of each other.

In summary, we have established that a) the enhanced mobility of at the surface of a glass-forming atomic alloy is a consequence of collective behaviour of the surface layers, b) the transverse component of the Debye-Waller factor of the interfacial inherent structure exhibits a sharp peak at the surface and c) the diffusion coefficient in the bulk and surface (transverse) share the same dependence on the value of the plateau $<\Delta r^2>$. Finally, we have demonstrated that the enhancement of the Debye-Waller factor that we calculate at the surface in the harmonic approximation accounts, when substituted into our general relation between the diffusion coefficient and the Debye-Waller factor, for a magnitude of surface enhanced mobility quite similar to that which has been reported from experiments. We conclude that the decrease in the transverse constraint at the surface, as measured by the inverse Debye-Waller factor, is *sufficient* to account for much of the kinetic enhancement. Should a longer correlation length develop at lower temperatures then it is possible that there will be an additional contribution to the surface kinetics, such as that proposed by Stevenson and Wolynes [25].

The value of connecting structure and dynamics is that the former is more accessible to theoretical and computational study than the latter, especially at low temperatures. The list of interesting questions about the surface dynamics of glasses is a long one. We have already raised the influence of curvature on surface dynamics. Other questions remain: what role does surface roughness play? Compositional ordering at the interface? What is the relationship between the transverse dynamics of a surface in 3D and the dynamics of 2D liquid? In establishing a connection between interfacial dynamics and the local Debye-Waller factor, we have identified a property obtainable from a single configuration, allowing us to explore these questions down to temperatures at which the dynamics themselves is no longer accessible to computer modelling.

**Acknowledgements**

We acknowledge support from the Discovery Program of the Australian Research Council and generous grants of computer time from High Performance Computing Facility at the University of Sydney.